\documentclass[%
 reprint,
 amsmath,amssymb,
 aps,
]{revtex4-2}

\usepackage{graphicx}
\usepackage{dcolumn}
\usepackage{bm}
\usepackage{hyperref}

\usepackage{algorithm}
\usepackage{algpseudocode}
\floatname{algorithm}{Metaheuristic:}
\makeatletter
\edef\ftype@algorithm{\number\ftype@algorithm}
\newcommand{\floatc@algdescription}[2]{{\small #2\par}}
\newcommand{\fs@ruledbottom}{%
  \fs@ruled
  \let\@fs@capt\floatc@algdescription
  \let\@fs@iftopcapt\iffalse}
\newcommand{\algorithmheading}{%
  \noindent\textbf{\fname@algorithm\ \thealgorithm}\par
  \kern2pt\hrule\kern2pt}
\makeatother
\floatstyle{ruledbottom}
\restylefloat{algorithm}
\usepackage{amsmath}

\usepackage{xcolor}
\usepackage[normalem]{ulem}

\begin{document}

\preprint{APS/123-QED}

\title{Competitive scaling of small-subsystem-optimization heuristics for LABS}

\author{Ginger E. Lau}
 \email{ginger.lau@alumni.emory.edu}
\affiliation{Physics \& Informatics Laboratories, NTT Research, Inc., Sunnyvale, California 94085, USA}

\author{Adil A. Gangat}%
 \email{aag38@caltech.edu}
\affiliation{Physics \& Informatics Laboratories, NTT Research, Inc., Sunnyvale, California 94085, USA}
\affiliation{Division of Chemistry and Chemical Engineering, California Institute of Technology, Pasadena, CA 91125}


\date{\today}

\begin{abstract}
The low autocorrelation binary sequences (LABS) problem, often formulated as a spin-lattice energy-minimization problem, is a combinatorial optimization problem with a very rugged energy landscape. We demonstrate that iterative optimization of very small subsystems of $n$ spins that are decoupled from the rest of the system achieves global optimization of LABS with a time-to-solution scaling that is competitive with heuristics that optimize the global system of $N$ spins directly.  Our isolated-subsystem heuristic with $n=4$ achieves a time scaling of $\mathcal{O}(1.33^N)$ on the interval $N=[27,60]$, while the current state-of-the-art classical and quantum scalings are respectively $\mathcal{O}(1.34^N)$ and $\mathcal{O}(1.21^N)$.  We surprisingly find that doubling the subsystem size to $n=8$ does not improve the scaling.  Our results suggest that specialized optimization hardware with only small numbers of logical spins may be of practical relevance, and that quantum speedups demonstrated in the absence of decomposition may not provide a practical quantum advantage for large-scale problems that require decomposition.

\end{abstract}

\maketitle


\section{\label{sec:intro}Introduction}
Specialized hardware, such as quantum annealers and coherent Ising machines, are under active development to solve combinatorial optimization problems with greater speed and less power than conventional CPUs and GPUs \cite{Johnson2011QuantumAnnealing,Honjo2021CIM}.  Since the number of optimization variables a device can accommodate is limited by engineering challenges, active investigation also aims to develop strategies for solving extremely large problems by decomposing them into subproblems composed of much smaller numbers of variables \cite{rosenberg2016building,Booth2017qbsolv}.  Here we investigate the time-scaling performance of one such decomposition strategy in the context of the low-autocorrelation binary sequences (LABS) problem, and we find evidence that it is competitive with the time scaling of methods that solve the LABS problem without such decompositions.

The LABS problem \cite{boehmer1967binary, schroeder1970synthesis}, which can be formulated as a spin-lattice energy-minimization problem, aims to determine the optimal sequence of binary variables which contains the lowest correlation to itself when shifted.
Motivations to determine optimal solutions are twofold. Due to the hardness of the problem (it contains a large number of degeneracies, which results in a very rugged energy landscape \cite{cadavid2025scaling}), it serves as a benchmark for the development of heuristics and metaheuristics \cite{gallardo2009finding}, particularly as a setting for quantum algorithms \cite{shaydulin2024evidence, sciorilli2025competitive, cadavid2025scaling} to demonstrate quantum speedup (i.e., a quantum time-to-solution scaling superior to the best known classical scaling). The solution sequences themselves are also of interest to signal processing and communications engineering \cite{boehmer1967binary, golay1977sieves}, where the key property that the optimal sequence has minimal correlation to itself is useful for signal clarity.

The LABS problem becomes intractable even at small system sizes. The known optimal LABS sequences up to length $N=48$ were originally obtained via exhaustive search \cite{mertens1996exhaustive}. Optimal sequences up to $N=66$ and optimal skewsymmetric sequences up to $N=119$ were obtained later \cite{packebusch2016low}, and excited energy levels have been analytically determined \cite{ukil2010low}.

Besides exact solution methods, a variety of heuristics have been developed, capable of estimating optimal LABS sequences up to larger $N$, but with no guarantee of optimality. These include classical metaheuristics, such as memetic tabu search \cite{glover1989tabu}, and quantum methods, such as the quantum approximate optimization algorithm (QAOA) \cite{farhi2014quantum}. 
The leading classical time scaling is memetic tabu search with $\mathcal{O}(1.34^N)$ \cite{gallardo2009finding, bovskovic2017low}, and the state-of-the-art quantum time scaling of  $\mathcal{O}(1.21^N)$ was inferred for QAOA enhanced with quantum minimum finding \cite{shaydulin2024evidence}.

Studies of the performance of decomposition-based metaheuristics on the LABS problem appear to be absent in the literature. Such heuristics optimize individual subsystems to find the global ground state \cite{dantzig1960decomposition, bertsekas1997nonlinear, rosenberg2016building}. Decomposition-based heuristics that work for dense-graph problems (like LABS) can be broadly divided into two categories considering how they treat the connections of a given subsystem to its environment before the subsystem is optimized: (1) methods that approximately incorporate the environment variables such that they become effective-field terms for the subsystem during its optimization \cite{Booth2017qbsolv, rosenberg2016building, Zhang2026ExtremeScale} and (2) methods that delete the interactions between the subsystem and its environment during the subsystem's optimization \cite{Bass2021Decomposition, Jo2022DivideConquer}.  In this work we devise and test heuristics for the LABS problem that belong to category (2). 

The heuristics we present in this paper consist of two parts. The first part concerns optimizations of chosen isolated subsystems of $n$ spins, which produces $n$-spin update proposals for the global system.  The subsystem optimization can be done with any method (we use exact enumeration in our simulations), so we refer to our heuristics as \textit{meta}heuristics. The second part consists of rules on how to accept or reject these $n$-spin update proposals.  We incorporate acceptance rules that consider the effect on global energy. Thus, although our metaheuristics discard information about the subsystem's environment from the first part, that information is implicitly used in the second part.

We are able to show that such isolated-subsystem-optimization metaheuristics can obtain the ground state solution for LABS sequences up to at least $N=60$. Remarkably, with a subsystem size of only $n=4$ we find a best runtime scaling of $\mathcal{O}(1.33^N)$, competitive with the classical state of the art $\mathcal{O}(1.34^N)$ from the full-system heuristic memetic tabu search.  We also surprisingly find that using $n=8$ yields slightly poorer performance than $n=4$.

The rest of the paper is structured as follows: We specify the LABS problem in Sec. \ref{sec:labs}.  We detail our metaheuristics in Sec. \ref{sec:metaheuristics}. We present simulation results of our metaheuristics in Sec. \ref{sec:simulations}.  And lastly, we summarize our findings and comment on their implications in Sec. \ref{sec:conclusion}.

\section{\label{sec:labs}The LABS problem}

For a LABS problem size $N$, the sequence of binary variables can be represented as
$ S = \{s_1, s_2, \dots, s_N\}$, where $s_i = \pm 1 $. The autocorrelation for an offset of $k$ spaces is:
\begin{equation}
    C_k(S) = \sum_{i=1}^{N-k} s_i s_{i+k},
\end{equation}
and the problem includes offset sizes $k=1, 2, \dots, N-1$. The solution is the sequence which minimizes the ``sidelobe energy":
\begin{equation}
    E(S) = \sum_{k=1}^{N-1} C_k^2(S)
    \label{eq:labs_energy}
\end{equation}
Mapping $S$ to Ising spins allows us to consider $E(S)$ as the Hamiltonian of a spin-lattice with two- and four-body interactions. This is the Bernasconi model \cite{bernasconi1987low, mertens1998ground} of statistical mechanics.

There exist three symmetry classes of the solution sequences. The energy is invariant under: reversal of the sequence $s_i \rightarrow s_{N+1-i}$, taking the complement $s_i \rightarrow -s_i$, and the alternate complement $s_i \rightarrow (-1)^i s_i$ where every other spin is the opposite sign of the existing sequence.

\section{\label{sec:metaheuristics}Metaheuristic methods}

\subsection{\label{sec:sec:base}Base metaheuristic}
The procedure of our simplest metaheuristic, which serves as a base for our other metaheuristics, is to optimize an isolated subsystem of $n$ spins of the $N$ total spins, then  update the $n$ spins in the full sequence. At each step, a different subsystem of $n$ spins is optimized and replaced in the full sequence, until the full optimal LABS sequence is obtained. We form the Hamiltonian used for the subsystem optimizations by discarding all terms in the global Hamiltonian that contain \textit{any} non-subsystem spins.

The \ref{alg:base} pseudocode below outlines the repeated steps described above. The function \textsc{RandomChoose}$(n)$ returns an $n$ number of spin indices to form the subsystem. The subsystem spin indices are then referred to as $s$. In the next step, the subsystem optimization is performed. The output of this step, $seq$, is an assignment of $1$ or $-1$ for each spin index in $s$. Finally, the $s$ subsystem spins in the full spin configuration are updated with the corresponding values obtained in $seq$. 

There is no guarantee that such updates will reduce the global energy.  This can be good for escaping \textit{local} minima, but can also result in moving away from \textit{global} minima. Therefore, part of what follows involves modifications to \ref{alg:base} such that global-energy-increasing updates are only sometimes accepted.

\begin{algorithm}
\renewcommand{\thealgorithm}{IsoSubsys$\_$Base}
\algorithmheading
\begin{algorithmic}[1]
\Require initial spin configuration

\Repeat

\State $s \gets$ \textsc{RandomChoose}$(n)$ 
\Comment{select $n$ spins}

\State $seq \gets$ Subsystem optimization of $s$
\State Update $s$ spins with $seq$
\Until{ground state energy is obtained}

\end{algorithmic}
\caption{ }
\label{alg:base}
\end{algorithm}

\subsection{Improvement features}

We consider four ways to improve upon \ref{alg:base} in the following four sections:

\begin{itemize}
    \item Sec.~\ref{sec:sec:sec:complement}: Testing subsystem complement
    \item Sec.~\ref{sec:sec:sec:kicked_descent}: Kicked descent
    \item Sec.~\ref{sec:sec:sec:choosers}: Selectively choosing subsystems
    \item Sec.~\ref{sec:sec:sec:multiple_Y}: Multiple subsystem proposals
\end{itemize}

We present three new metaheuristics (\ref{alg:SC}, \ref{alg:KD}, and \ref{alg:MS}) by sequentially adding the four improvements to \ref{alg:base}. Adding a test of subsystem complements to the base metaheuristic results in \ref{alg:SC}.  Subsequently adding a procedure that searches for energy-lowering updates while occasionally permitting energy-raising updates (i.e., ``kicked descent") results in \ref{alg:KD}.  Adding to that a systematic way of choosing subsystems and also a way of considering multiple subsystem-update proposals at a time results in \ref{alg:MS}. Table~\ref{tab:algs_comp} summarizes the features of these improved metaheuristics.

\subsubsection{\label{sec:sec:sec:complement}Testing subsystem complement}

The first feature expands upon \ref{alg:base} by considering both the obtained subsystem spin predictions and also its complement (multiply by $-1$). This is incorporated in \ref{alg:SC}.  While the sequence and its complement have identical subsystem energy due to the symmetry, when replacing the subsystem spins in the full LABS sequence, they can result in different \textit{global} energies. This process is indicated by \textsc{TestSequence}$(seq1, seq2)$ in the pseudocode. This feature calculates the global energy of using each of $seq1$ and $seq2$ in the full-$N$ sequence, and returns the lower global energy of the two. \ref{alg:SC} then updates the full-$N$ sequence with whichever of $seq1$ and $seq2$ gives the lower global energy.  Appendix~\ref{sec:app:posneg} provides data that shows how testing the subsystem complement substantially improves the performance of \ref{alg:SC} compared to \ref{alg:base}.

\begin{algorithm}
\renewcommand{\thealgorithm}{IsoSubsys$\_$SC}
\algorithmheading
\begin{algorithmic}[1]
\Require initial spin configuration 

\Repeat
\State $s \gets $ \textsc{RandomChoose}(n)
\State $seq1 \gets$ Subsystem optimization of $s$

\State \textcolor{red}{$seq2 \gets -seq1$ }
\State \textcolor{red}{ $E_{new} \gets$ \textsc{TestSequence}$(seq1, seq2)$ } 
\textcolor{red}{\Comment{Returns lower global energy} }
\State Update $s$ spins with $seq1$ or $seq2$, determined by $E_{new}$
\Until{ground state energy is obtained}

\end{algorithmic}
\caption{Adds a test of the subsystem complement to \ref{alg:base}}
\label{alg:SC}
\end{algorithm}

\subsubsection{\label{sec:sec:sec:kicked_descent}Kicked Descent}

The ``Kicked Descent" improvement results in the \ref{alg:KD} metaheuristic.  It entails being selective about which proposed updates are accepted. \ref{alg:KD} tests both the subsystem sequence and its complement, and then updates according to the one that lowers global energy the most, and does not update at all if neither choice lowers the global energy.

However, in the rugged energy landscape of the LABS problem, this procedure quickly traps the search in a local minimum. Therefore, \ref{alg:KD} also occasionally runs one cycle of \ref{alg:base}: it selects a random subset of $n$ spins and always accepts the update, regardless of how it changes the global energy. The hyperparameter controlling this process, $X$, is defined as the number of consecutive failed attempts to lower the global energy that must occur before employing one step of \ref{alg:base} to potentially kick the system out of the current local minimum.  Simulations in Sec. \ref{sec:simulations} show that \ref{alg:KD} has substantially better scaling and time-to-solution than \ref{alg:SC}.

\begin{algorithm}
\renewcommand{\thealgorithm}{IsoSubsys$\_$KD}
\algorithmheading
\begin{algorithmic}[1]
\Require initial spin configuration, \textcolor{red}{threshold $X$ }
\State \textcolor{red}{$x \gets 0$ }

\Repeat
\If{\textcolor{red}{$x < X$} }
\State $s \gets \textsc{RandomChoose}(n)$ 

\State $seq1 \gets$ Subsystem optimization of $s$
\State $seq2 \gets -seq1$

\State $E_{new} \gets$ \textsc{TestSequence}$(seq1, seq2)$ \Comment{Returns lower energy}

\If{ \textcolor{red}{$E_{new} < E_{current}$ } }
    \State Update $s$ spins with $seq1$ or $seq2$, determined by $E_{new}$
    \State \textcolor{red}{$x \gets 0$ }
\Else 
    \State \textcolor{red}{$x \gets x + 1$ }
\EndIf

{\algrenewcommand\algorithmicelse{\textcolor{red}{\textbf{else}}} \Else }
\Comment{ \textcolor{red}{One step of Base when $x = X$} }
\State $s \gets$ \textsc{RandomChoose}$(n)$ 

\State $seq \gets$ Subsystem optimization of $s$
\State Update $s$ spins with $seq$
\State \textcolor{red}{$x \gets 0$ }

\EndIf

\Until{ground state energy is obtained}

\end{algorithmic}
\caption{Adds a ``kicked descent" to \ref{alg:SC}: looks for energy-lowering updates, but allows an energy-raising update after $X$ failures to lower energy.}
\label{alg:KD}
\end{algorithm}

\subsubsection{\label{sec:sec:sec:choosers}Selectively choosing subsystems}

In \ref{alg:base}, \ref{alg:SC}, and \ref{alg:KD}, the choice of subsystem indices are randomly drawn from the $N$ spins at each call (``\textsc{RandomChoose}" in the pseudocode). Here we introduce a more sophisticated version of the subsystem index selection method, named \textsc{TrackedChoose} in the pseudocode.

\textsc{TrackedChoose} tracks which spins have been selected for subsystems and prevents them from being chosen for subsequent subsystems, with the exception of one spin that is included in each subsystem, until all $N$ spins have been selected to a subsystem.  If less than $n$ spins remain unselected during such a cycle, then the last subsystem in the cycle will consist of however many spins remain.  The single spin that is chosen as the overlapping spin between all the subsystems is allowed to change its value with each subsystem optimization.  The motivation for having this overlap feature is to provide a minimal way of unfreezing previously frozen spins in a way that could aid the global optimization.  As an example with $n=7$, the first call to \textsc{TrackedChoose} will return seven different random spins, the second call will return one spin from the first set of seven plus six spins randomly chosen from other than the first seven, and so on until all $N$ spins have been chosen for a subset.  If there is a remainder of less than six unselected spins at the end of this selection process, the final subset is formed by including the remaining selected spins plus the single overlap spin for all the subsets.


\subsubsection{\label{sec:sec:sec:multiple_Y}Multiple subsystem proposals}
The final improvement feature allows multiple subsystem updates to be proposed for a single update, and the one that produces the lowest global energy is chosen for the update.  While this entails multiple cost-function evaluations per update, it may reduce the number of updates needed to reach the global minimum and thereby still yield a scaling improvement.  We name the new hyperparameter controlling this $Y$, and it dictates the total number of proposed subsystem updates that are considered for a single actual update. Pseudocode for the \ref{alg:MS} metaheuristic shows how \textsc{TrackedChoose} and multiple subsystem proposals are incorporated into \ref{alg:KD}.

\begin{algorithm}
\renewcommand{\thealgorithm}{IsoSubsys$\_$MS}
\algorithmheading
\begin{algorithmic}[1]
\Require initial spin configuration, threshold $X$, \textcolor{red}{number of subsystem candidates $Y$ }

\State $x \gets 0$ 

\Repeat
\If{$x < X$} 

\For{ \textcolor{red}{$y = 1,\dots,Y$} }

\State \textcolor{red}{ $s \gets \textsc{TrackedChoose}(n)$} 

\State $seq1 \gets$ Subsystem optimization of $s$
\State $seq2 \gets -seq1$

\State $E_{new} \gets$ \textsc{TestSequence}$(seq1, seq2)$

\If{ \textcolor{red}{$E_{new} < E_{current}$ } }
    \State \textcolor{red}{$seq_{best} \gets seq1$ or $seq2$, determined by $E_{new}$}
\EndIf

\EndFor

\If{ \textcolor{red}{$seq_{best}$} }
    \State \textcolor{red}{Update $s$ spins with $seq_{best}$ }
    \State $x \gets 0$
\Else 
    \State $x \gets x + 1$ 
\EndIf

\Else
\Comment{One step of Base when $x = X$ }
\State $s \gets$ \textsc{RandomChoose}$(n)$ 

\State $seq \gets$ Subsystem optimization of $s$
\State Update $s$ spins with $seq$
\State $x \gets 0$ 

\EndIf

\Until{ground state energy is obtained}

\end{algorithmic}
\caption{Modifies \ref{alg:KD} with \textsc{TrackedChoose} and a test of multiple distinct subsystems per update.}
\label{alg:MS}
\end{algorithm}

\begin{table*}
\caption{\label{tab:algs_comp}
Metaheuristic features and hyperparameters.  Hyperparameter choices are from the tuning described in Sec. \ref{sec:sec:tuning}.}

    \centering
    \begin{ruledtabular}
    \begin{tabular}{ccccc}
        Metaheuristic & Subsys. Complement Check? & $X$ & Subsys. Selector & $Y$ \\
        \hline
        IsoSubsys$\_$Base & No & -  & \textsc{Random} & - \\
        
        IsoSubsys$\_$SC & Yes & -  & \textsc{Random} & - \\

        IsoSubsys$\_$KD & Yes & 80 & \textsc{Random} & -  \\

        IsoSubsys$\_$MS ($n=4$) & Yes & 90 & \textsc{Tracked} & 50 \\
        IsoSubsys$\_$MS ($n=8$) & Yes & 70 & \textsc{Tracked} & 45 \\

    \end{tabular}
    \end{ruledtabular}
\end{table*}

\section{Simulations}
\label{sec:simulations}

\subsection{TTS definition and fitting}
\label{sec:TTS}
We define time-to-solution (TTS) as the total number of evaluations of the full $N$ sequence for global energy.  This definition of TTS is to allow for a sensible comparison with recent literature results on TTS scaling of heuristics in the context of LABS: Sciorilli \textit{et al.} \cite{sciorilli2025competitive} and Cadavid \textit{et al.} \cite{cadavid2025scaling} both use this TTS definition, and Shaydulin \textit{et al.} \cite{shaydulin2024evidence} uses the expected number of measurements required to reach an optimal solution with QAOA.

In \ref{alg:SC}, \ref{alg:KD}, and \ref{alg:MS}, each subsystem selected contributes two units of TTS: one for testing the subsystem update proposal and one for testing its complement.

As in the recent literature, we fit the exponential curve
\begin{equation}
    \textrm{TTS} = c \cdot b^N
\label{eq:TTS}
\end{equation}
by performing linear regression on $\log(\textrm{TTS})$. The fits are to the median TTS for each problem size $N$.  Each metaheuristic is tested on the same 20 seeds per $N$, with each seed corresponding to an initial global spin configuration. 

The system sizes $N$ that are studied in the recent literature differ: Shaydulin \textit{et al.} uses $N=[28,40]$ and Cadavid \textit{et al.} uses $N=[27,37]$. Sciorilli \textit{et al.} splits into separate fits for odd and even $N$, using $N=20, 24, 28, 32, 34, 36, 38, 40, 42, 44$ for even and $N=13, 21, 27, 41, 43 $ for odd.  We report the results of our metaheuristics on all these sets of $N$, as well as the set $N=[27, 60]$.

\subsection{Hyperparameter tuning}
\label{sec:sec:tuning}
As detailed in Sec. \ref{sec:metaheuristics}, the metaheuristic \ref{alg:KD} contains a hyperparameter $X$, and the metaheuristic \ref{alg:MS} contains hyperparameters $X$ and $Y$. These hyperparameters are tuned to minimize the exponential base $b$ in Eq. (\ref{eq:TTS}) on the interval $N=[27, 50]$, and they are tuned independently for \ref{alg:KD} and \ref{alg:MS}, and also for each subsystem size $n$. The same ten seeds are used for both tunings. Table~\ref{tab:algs_comp} gives the tuned values that are used to generate the results reported below, and further details of the tunings are given in Appendix \ref{sec:app:hyperparam}.

\subsection{Results}
\begin{figure}[tb]
    \centering
    \includegraphics[width=\columnwidth]{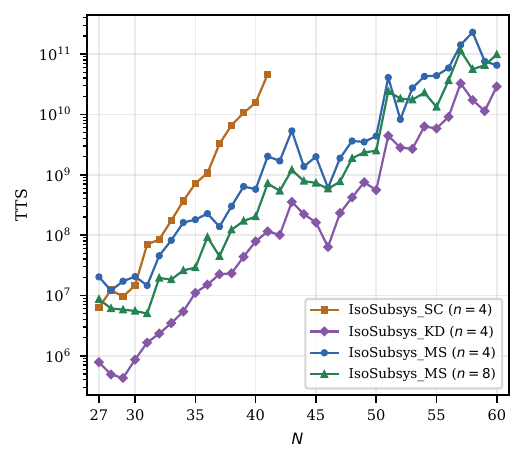}
    \caption{Time-to-solution (TTS), measured as the median number of global energy evaluations required to reach the known LABS optimum over twenty seeds at each $N$.}
    \label{fig:TTS}
\end{figure}

\begin{figure}[h]
    \centering
    \includegraphics[width=\columnwidth]{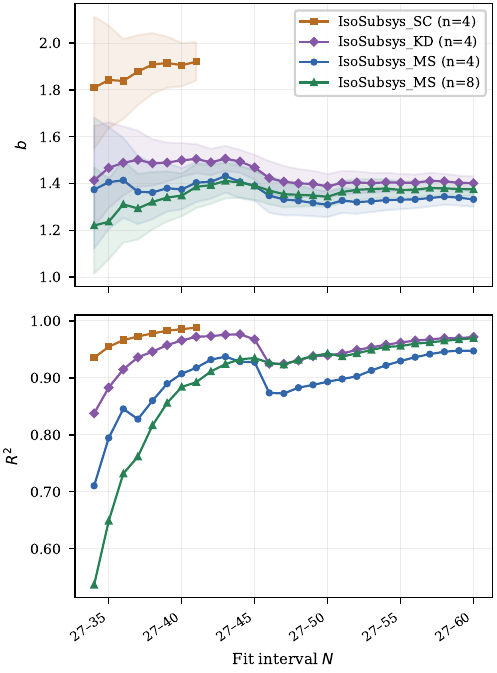}
    \caption{Fitted scaling base $b$ (top) and its log-space $R^2$ (bottom) versus fit intervals. Fits are to the per-$N$ median number of global energy evaluations over twenty uncapped seeds. Shading: 95\% Student-$t$ confidence intervals.}
    \label{fig:scaling}
\end{figure}

\begin{figure}[tb]
    \centering
    \includegraphics[width=\columnwidth]{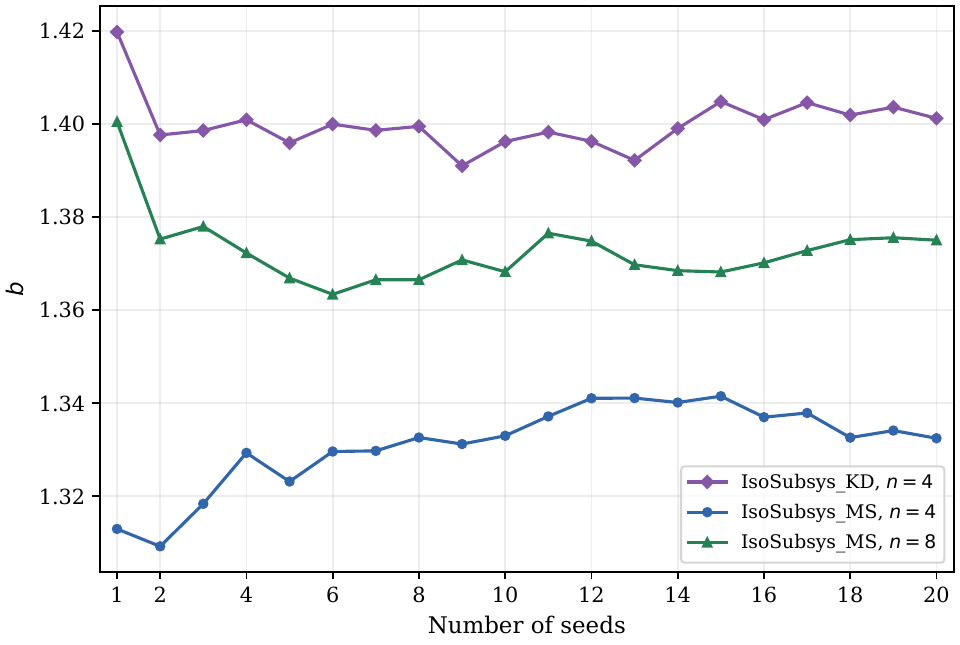}
    \caption{Fitted exponential base $b$ versus number of seeds for IsoSubsys\_KD ($n=4$) and IsoSubsys\_MS ($n=4,8$), over the interval $N=[27, 60]$. The subsets are nested and follow seed-label order; points are not averages over random seed subsets.}
    \label{fig:b_vs_seeds}
\end{figure}

As shown in Appendix \ref{sec:app:posneg}, \ref{alg:base} becomes vastly slower than \ref{alg:KD} beyond $N=30$ for both $n=4$ and $n=8$, so we do not assess its performance further here. In Fig. \ref{fig:TTS} we present TTS results for the improved metaheuristics with $n=4$, and additionally with $n=8$ only for \ref{alg:MS}.  Scaling results from this data are presented in Fig. \ref{fig:scaling}.  While \ref{alg:KD} has the smallest TTS over the values of $N$ tested, \ref{alg:MS} has slightly better scaling. Curiously, doubling $n$ from four to eight slightly worsens the performance of \ref{alg:MS} in terms of both the TTS and its scaling. In Fig. \ref{fig:b_vs_seeds} we show that $b$ for our metaheuristics on the interval $N=[27, 60]$ is converged in the number of seeds to about $1\times10^{-2}$.

Table~\ref{tab:scaling_comp_best} compares the scaling performance of \ref{alg:KD} and \ref{alg:MS} with the results in the recent literature and also reports our results on the full interval $N=[27, 60]$. We note that the scaling of \ref{alg:MS} with $n=4$ on $N=[27, 60]$ is competitive with the current classical state-of-the-art from memetic tabu \cite{shaydulin2024evidence, gallardo2009finding, bovskovic2017low}.  The trend in Fig. \ref{fig:scaling} suggests that this competitive scaling will persist as the upper bound on the interval of $N$ grows beyond what we have checked, especially since $N=60$ is in the $N\gg n$ limit for $n=4$.

\begin{table}[tb]
    \caption{\label{tab:scaling_comp_best}
    Fits of Eq. (\ref{eq:TTS}) for IsoSubsys$\_$KD and IsoSubsys$\_$MS compared with reported literature results. Fits are to medians of twenty uncapped seeds per $N$. ``Odd'' and ``Even'' are what is specified at the end of Sec. \ref{sec:TTS}.}
    \centering
    \footnotesize
    \setlength{\tabcolsep}{2pt}
    \begin{ruledtabular}
    \begin{tabular}{ccccc}
        Heuristic & $N$ & $b$ & $CI$ & $R^2$ \\
        \hline
        QE-MTS \cite{cadavid2025scaling} & [27, 37] & $\approx 1.24$ & 1.23--1.25 & [0.86,0.89]$^{a}$ \\
        IsoSubsys$\_$KD ($n=4$) & [27, 37] & 1.466 & 1.333--1.613 & 0.9019 \\
        IsoSubsys$\_$MS ($n=4$) & [27, 37] & 1.427 & 1.251--1.627 & 0.8065 \\
        IsoSubsys$\_$MS ($n=8$) & [27, 37] & 1.293 & 1.160--1.442 & 0.7617 \\
        \hline
        QAOA+QMF \cite{shaydulin2024evidence} & [28, 40] & 1.21 & 1.19--1.23 & $>0.94$ \\
        QAOA \cite{shaydulin2024evidence} & [28, 40] & 1.46 & 1.42--1.50 & $>0.94$ \\
        Memetic tabu \cite{shaydulin2024evidence} & [28, 40] & 1.35 & 1.33--1.38 & NR \\
        Branch \& bound \cite{shaydulin2024evidence} & [28, 40] & 1.615 & 1.571--1.659 & NR \\
        IsoSubsys$\_$KD ($n=4$) & [28, 40] & 1.510 & 1.417--1.609 & 0.9488 \\
        IsoSubsys$\_$MS ($n=4$) & [28, 40] & 1.433 & 1.320--1.555 & 0.8948 \\
        IsoSubsys$\_$MS ($n=8$) & [28, 40] & 1.394 & 1.309--1.485 & 0.9250 \\
        \hline
        PCE \cite{sciorilli2025competitive} & Odd & 1.405 & 1.212--1.629 & 0.94 \\
        IsoSubsys$\_$KD ($n=4$) & Odd & 1.398 & 1.327--1.472 & 0.9930 \\
        IsoSubsys$\_$MS ($n=4$) & Odd & 1.391 & 1.342--1.442 & 0.9965 \\
        \hline
        PCE \cite{sciorilli2025competitive} & Even & 1.338 & 1.261--1.420 & 0.94 \\
        IsoSubsys$\_$KD ($n=4$) & Even & 1.427 & 1.370--1.486 & 0.9809 \\
        IsoSubsys$\_$MS ($n=4$) & Even & 1.358 & 1.260--1.464 & 0.9177 \\
        \hline
        IsoSubsys$\_$KD ($n=4$) & [27, 60] & 1.396 & 1.363--1.430 & 0.9621 \\
        IsoSubsys$\_$MS ($n=4$) & [27, 60] & 1.333 & 1.300--1.366 & 0.9462 \\
        IsoSubsys$\_$MS ($n=8$) & [27, 60] & 1.375 & 1.347--1.403 & 0.9697 \\
    \end{tabular}
    \end{ruledtabular}
    \par\smallskip
    \parbox{\columnwidth}{\footnotesize\raggedright
    CIs are 95\% Student-$t$ intervals from log-linear regression; $R^2$ is on the log scale. NR means not reported.
    $^{a}$QE-MTS: the last entry is their 95\% bootstrap CI for $R^2$.
    The memetic tabu and branch-and-bound fits in Ref.~\cite{shaydulin2024evidence} are to mean CPU time, with branch-and-bound stopped at the first optimal solution.}
\end{table}

\section{Summary and outlook}
\label{sec:conclusion}
Solving very-large-scale discrete-variable optimization problems on quantum computers and/or specialized classical hardware will likely, for the foreseeable future, require decomposing the problems into much smaller sizes that can fit on the hardware.  This work investigates, in the context of the LABS problem, how the time-scaling performance of optimization heuristics that work with such decomposed problems compares to the performance of state-of-the-art heuristics that attack problems as a whole.

This work provides evidence for two surprising results: (1) decomposition-based heuristics that use very small subsystems of only four variables yield time-scaling performance on LABS that is competitive with classical full-system-optimization heuristics, and (2) increasing the subsystem size to eight variables does not increase the performance of the heuristics.  If future work on LABS and other problems generalize and otherwise strengthen these results, the implications for hardware design could be very substantial: highly efficient hardware with only a few logical variables may be an overall better proposal than much larger hardware.  In other words, the results in this paper raise the following question: when, if ever, do hardware solvers with large numbers of variables substantially outperform hardware solvers with only a few variables on extremely large combinatorial optimization problems?  A practical approach to this question would need to take into account walltime and power consumption in addition to time scaling.  We conjecture that small-scale hardware yields better performance for problems with dense interaction graphs, while large-scale hardware outperforms on sparse-graph problems; this would be consistent with the result in this paper, which deals with a dense-graph problem, that doubling the subsystem size from four to eight yielded slightly worse time-scaling on the largest range of system sizes fitted.

Our results also relate to the topic of quantum advantage: they cast some doubt on the practical relevance of quantum speedups that belong to quantum algorithms that do not invoke problem decomposition (e.g., Ref. \cite{shaydulin2024evidence, boulebnane2024solving}). As mentioned, for very-large-scale problems, decomposition will be necessary. Thus, for quantum speedups belonging to non-decomposition quantum algorithms to be of benefit in the context of decomposition, they must provide subsystem optimizations that are faster than any classical method. However, this is only likely to happen at subsystem scales beyond a few variables, while our results suggest (but do not prove) that increasing subsystem sizes beyond a few variables does not improve the time-to-solution scaling for the full problem when it is a dense-graph problem.

Regarding our metaheuristics themselves, there are various ways left open to investigate how to further improve their time scaling and walltime. For example, decomposition methods that treat the subsystem boundaries differently than the simple disconnection approach used in this paper may yield improved performance.

Finally, it is noteworthy that the best scaling we achieve with our small-subsystem-optimization metaheuristics is roughly the same as the currently-known best scaling of classical full-system-optimization heuristics.  Is this a mere coincidence, or is it indicative of something fundamental?  Attempts to improve the scaling of our metaheuristics could shed light on this question.

\begin{acknowledgments}
The authors acknowledge the use of OpenAI Codex for assistance with implementing and managing numerical simulations, analyzing data, preparing figures, and editing the manuscript.
\end{acknowledgments}

\appendix

\section{\label{sec:app:posneg}Benefits of testing sequence complement} 

\begin{figure*}[t]
    \centering
    \includegraphics[width=\textwidth]{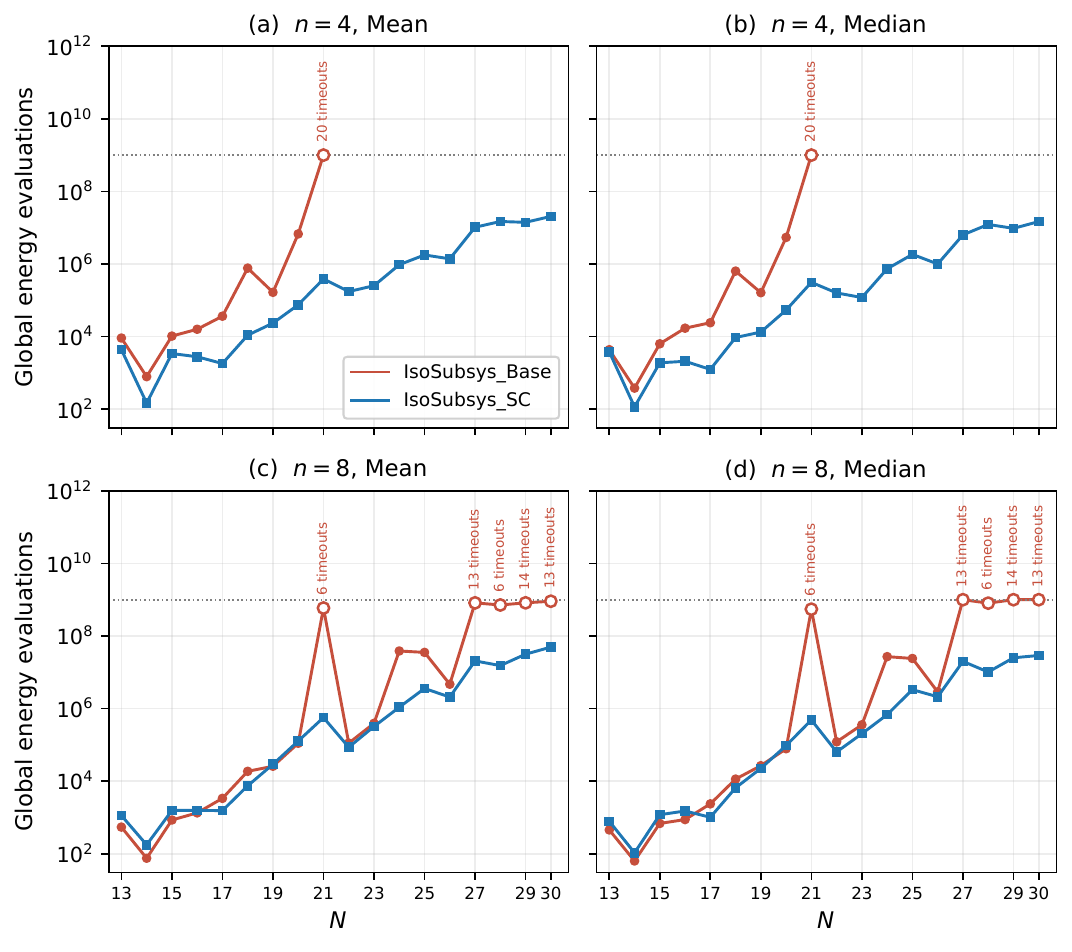}
    \caption{Global energy evaluations to reach the known LABS optimum for IsoSubsys\_Base and IsoSubsys\_SC, using exact optimization of isolated subsystems of size $n=4$ (top) and $n=8$ (bottom). Left: mean; right: median over twenty matched seeds per $N$. Each seed is limited to $10^9$ global energy evaluations, with no walltime limit for new runs. Open circles show the mean or median including budget-exhausted seeds at their observed capped counts, so these statistics are lower bounds on uncapped values; annotations give the number of exhausted seeds. The dotted line marks the call cap.}
    \label{fig:A:Base_vs_SC}
\end{figure*}

In our metaheuristics, a given subsystem is optimized and the resulting subsystem configuration is used as an update proposal.  A key finding of this study is that testing both this subsystem-optimization-proposed update as well as its complement significantly lowers runtime to reach the solution. We illustrate this with a comparison of \ref{alg:base} and \ref{alg:SC} in Fig.~\ref{fig:A:Base_vs_SC}.

\section{\label{sec:app:hyperparam}Hyperparameters X and Y}

Here we provide details of the tuning of hyperparameters $X$ and $Y$ mentioned in Sec.~\ref{sec:sec:tuning}.

\subsection{Tuning objective and sampling}
For each tested combination of $X$ and $Y$, we use every integer $N=27,\ldots,50$ and the same ten seeds at each $N$. Each run continues until it reaches the known optimum, with no evaluation-count or walltime cap.

Let $M_N$ be the median number of global energy evaluations required to reach the optimum over the ten seeds at size $N$. We perform an unweighted ordinary least-squares fit
\begin{equation}
    \log M_N = a+sN, \qquad b=\exp(s),
\end{equation}
using all 24 medians, giving each system size equal weight. The candidate with the smallest fitted $b$ is selected. The coefficient of determination $R^2$ is calculated in log space and reports fit quality; it is not the primary tuning objective.

\subsection{Coarse and fine grid searches}
For \ref{alg:KD}, only $X$ is tuned; $Y=1$ is fixed. For \ref{alg:MS}, $X$ and $Y$ are tuned jointly. The initial coarse grid is
\begin{align}
    X&\in\{70,80,90,100,110\},\\
    Y&\in\{10,20,30,40\}\quad\text{(IsoSubsys\_MS only)}.
\end{align}
For \ref{alg:MS}, every combination of $X$ and $Y$ in the grid is evaluated. If the best point lies on a boundary, that boundary is extended in increments of 10 and the added settings are evaluated. After an interior coarse-grid minimum is found, a fine grid is formed from the selected value and its neighbors at $\pm5$ along each tuned axis. If the fine-grid minimum is on a boundary, that boundary is extended in increments of 5 until the minimum is surrounded (or the positive-integer boundary is reached).

The grids actually evaluated were as follows:
\begin{itemize}
    \item \textbf{IsoSubsys\_KD, $n=4$:} the initial five-point coarse $X$ grid required no expansion. The fine grid was $X\in\{75,80,85\}$. Seven distinct settings were evaluated in total, selecting $X=80$.
    \item \textbf{IsoSubsys\_MS, $n=4$:} the coarse $X$ axis remained unchanged, while the coarse $Y$ axis expanded to $\{10,20,30,40,50,60\}$. The fine grid was $X\in\{85,90,95\}$ and $Y\in\{45,50,55\}$. The 38 distinct settings selected $(X,Y)=(90,50)$.
    \item \textbf{IsoSubsys\_MS, $n=8$:} the coarse axes expanded to $X\in\{60,70,80,90,100,110\}$ and $Y\in\{10,20,30,40,50\}$. The initial fine grid was $X\in\{65,70,75\}$ and $Y\in\{35,40,45\}$; the $Y$ axis was then extended to 50. The 40 distinct settings selected $(X,Y)=(70,45)$.
\end{itemize}

\subsection{Dependence on the fitted system-size range}
\begin{figure}[tb]
    \centering
    \includegraphics[width=\columnwidth]{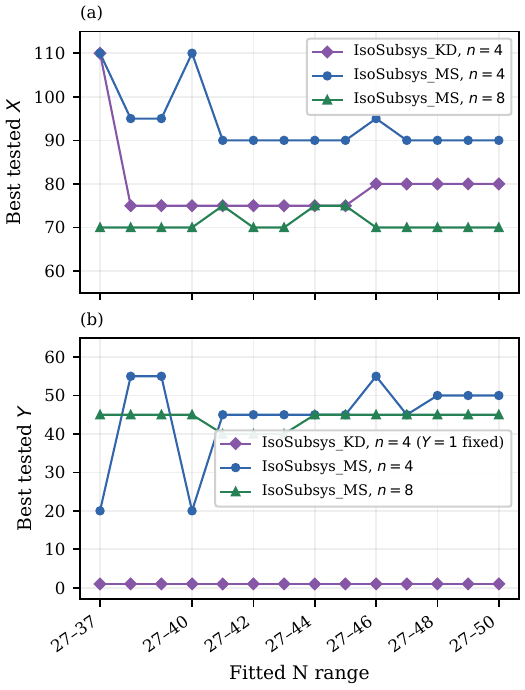}
    \caption{Best tested (a) $X$ and (b) $Y$ versus fitted $N$ range for IsoSubsys\_KD with $n=4$ and IsoSubsys\_MS with $n=4,8$. Each point minimizes the exponential base $b$ from an unweighted log-linear fit to per-size medians of ten seeds over every integer $N$ in the indicated interval. The fixed candidate sets contain 7, 38, and 40 settings, respectively. IsoSubsys\_KD has fixed $Y=1$. Lines connect discrete grid selections as visual guides. The final interval, $N=27$--50, reproduces the selected parameters in Table~\ref{tab:tuning_results}.}
    \label{fig:optimal_parameters_windows}
\end{figure}

Fig.~\ref{fig:optimal_parameters_windows} shows how the best tested hyperparameters depend on the fitted interval. We refit the saved ten-seed data on each expanding interval $N=27,\ldots,N_{\mathrm{upper}}$, with $N_{\mathrm{upper}}=37,\ldots,50$, and select the candidate with the smallest $b$. For each metaheuristic and subsystem size, the candidate set is held fixed at all settings evaluated in its completed coarse and fine searches described above. These are retrospective comparisons within those tested grids, not separate adaptive tuning searches or independently validated global optima. For IsoSubsys\_MS, both panels show coordinates of the same jointly selected $(X,Y)$ point; for IsoSubsys\_KD, $Y=1$ remains fixed. No additional simulations are used.

\subsection{Selected parameters and subsequent evaluation}
Table~\ref{tab:tuning_results} reports the fits at the selected settings. These are ten-seed tuning fits on $N=27$--50, distinct from the twenty-seed fits and other fitting intervals reported in the main text.
\begin{table}[tb]
    \caption{Selected parameters and ten-seed tuning fits over all integer $N=27$--50. The reported $R^2$ is in log space.}
    \label{tab:tuning_results}
    \centering
    \begin{ruledtabular}
    \begin{tabular}{lccccc}
        Metaheuristic & $n$ & $X$ & $Y$ & $b$ & $R^2$ \\
        \hline
        IsoSubsys\_KD & 4 & 80 & 1 & 1.361072 & 0.920050 \\
        IsoSubsys\_MS & 4 & 90 & 50 & 1.311048 & 0.892370 \\
        IsoSubsys\_MS & 8 & 70 & 45 & 1.330555 & 0.901892 \\
    \end{tabular}
    \end{ruledtabular}
\end{table}

\bibliography{refs}

\end{document}